\documentclass[aps,prl,floatfix,twocolumn,superscriptaddress,nofootinbib,a4paper]{revtex4-1}
\pdfoutput=1

\usepackage{amsmath}
\usepackage{amssymb}
\usepackage{graphicx}
\usepackage{hyperref}
\usepackage{color}
\usepackage{xcolor}
\usepackage[T1]{fontenc} 
\usepackage{slashed}
\usepackage{comment}
\usepackage{tikz}

\definecolor{myblue}{RGB}{79,129,189}
\colorlet{darkblue1}{myblue!130}
\colorlet{darkblue}{darkblue1!82!black}

\usepackage[normalem]{ulem} 

\definecolor{blus}{cmyk}{1,1,0,0.6}
\hypersetup{colorlinks,bookmarksopen,bookmarksnumbered,linkcolor=blus,pdfstartview=FitH,urlcolor=blus,citecolor=blus}

\allowdisplaybreaks

\newcommand{\AddrOXF}{%
Rudolf Peierls Centre for Theoretical Physics, University of Oxford, Parks Road, Oxford OX1 3PU, UK
}

\begin{document}

\title{A mass for the dual axion}

\author{Arthur Platschorre}\email{arthur.platschorre@physics.ox.ac.uk}
\affiliation{\AddrOXF}

\begin{abstract}\noindent
In this article we study a modification of axion physics in which the dual axion acquires a mass. This mass explicitly breaks the shift symmetry of the dual axion. The potential breaking of this shift symmetry poses a dual axion quality problem. When the dual axion acquires a mass, the axion gets eaten and becomes the longitudinal degree of freedom of a massive vector field. In this phase, axion strings are screened and far-separated instanton configurations are exponentially suppressed. This confinement of instantons corresponds to the worldline action of a particle-like soliton traveling between the instantons analogous to Abrikosov/Nielsen-Oleson vortex solitons that stretch between confined magnetic monopoles in a superconductor. We calculate the cost of this additional worldline suppression and provide several models in which both the confined instantons and confining worldline are dynamical. 
\end{abstract}

\maketitle

\section{Introduction}
Among the many beyond the Standard Model (BSM) extensions, the axion, a  (pseudo)-Nambu-Goldstone boson, has received ever-growing attention in recent decades. The QCD axion would provide a simple explanation of the strong CP problem \cite{Peccei:1977hh, Wilczek:1977pj,Weinberg:1977ma}, axions would provide an elegant solution to several cosmological questions such as the nature of dark matter \cite{Preskill:1982cy,Abbott:1982af,Dine:1982ah} or dark energy \cite{Marsh:2012nm,PhysRevD.83.123526} and axions are expected to be ubiquitous in string theory compactifications \cite{Svrcek:2006yi,Arvanitaki:2009fg}.  

Most of the desired properties of an axion rely on the existence of an (approximate) continuous shift symmetry, making the axion generically very light. In quantum gravity, 
such global symmetries are expected to be absent \cite{Banks:1988yz,Kallosh:1995hi,Banks:2010zn}. The axion's shift symmetry is expected to be either broken or gauged.

The breaking of the axion's shift symmetry by  sensitivities to UV physics and a potential loss of control over the axion mass provides a quality problem for parametrically light axions. Standard arguments for the QCD axion require such (misaligned) sources to have either extremely suppressed couplings or to be generated at a (very) high order in operators (for a recent review see \cite{Hook:2018dlk}). This has been dubbed the QCD axion quality problem \cite{Kamionkowski:1992mf,Dine:1986bg}.  

Recent investigations have re-examined the QCD axion quality problem from the perspective of the dual of the QCD axion \cite{Kallosh:1995hi,Quevedo:1995ep,Burgess:2023ifd,Dvali:2022fdv,Sakhelashvili:2021eid,Choi:2023gin}, the associated two-form Kalb-Ramond field \cite{Kalb:1974yc}. From this dual perspective, breaking the axion shift symmetry corresponds to a gauged shift symmetry of the dual axion. 

In this article, we consider the second option allowed by quantum gravity, a gauged shift symmetry of the axion. The axion gets `eaten' and becomes the longitudinal degree of freedom of a massive vector field. Generically, such axions arise in string theory in anomaly cancellation mechanisms in either compactifications of heterotic string theory \cite{Svrcek:2006yi} or type II intersecting brane models \cite{Ibanez:2001nd,Blumenhagen:2006ci,Antoniadis:2002qm}, resulting in massive vector fields with a stringy mass scale. 

In the presence of this massive vector field, the vacuum acts as a superconductor. Electric field lines can end in the vacuum and field lines emanating from electrically charged particles are screened at long distances. Magnetic field lines are forced into flux tubes and magnetic monopoles are confined. 

This motivates a general question: \textit{What happens to instantons and strings coupled to the axion in this superconducting phase?}

As the axion's degree of freedom is gauged, we turn to the dual description. From the dual perspective, a gauged shift symmetry of the axion corresponds to a mass for the dual axion. We will briefly review this dual perspective and potential mass generating mechanisms in Section \ref{section1}. A potential loss of control over the dual axion mass due to sensitivities to UV physics presents a dual axion quality problem and dramatically alters the interactions of the axion with axion strings and instantons. 

Strings and instantons coupled to the axion are respectively `electrically' and `magnetically' charged under the dual axion. Analogous to a massive vector field, a massive dual axion implies that the `electric flux' lines of the dual axion can end in the vacuum and the dual axion profile around strings is screened at large distances. The dual axion's `magnetic flux' lines surrounding instantons are forced into `flux tubes' and instantons are confined. This phase should be contrasted to a phase in which instead the axion is massive; strings are confined by axion domain walls and the axion field around instantons is screened at large distances.

These effects are a natural consequences of gauging the axion shift symmetry. The axion profile around axion strings is compensated by that of the vector field and the strings become non-interacting at large distances. Interactions with their environment can come from coupling to other long-range fields such as electromagnetism or gravity. Their phenomenology is that of gauge strings  and we connect back to this point of view in Section \ref{section2}.  

The gauged axion shift symmetry also implies that instantons coupled to the axion exist only in charge neutral dipole configurations and any far-separated instanton configurations are exponentially suppressed. The absence of isolated instantons in such theories was already noted in \cite{Heidenreich:2020pkc,Heidenreich:2021xpr}. In Section \ref{section3}, we show that this confinement of instantons corresponds to the worldline action of a particle-like soliton traveling between the instantons. This soliton is analogous to Abrikosov/Nielsen-Oleson vortex solitons that stretch between confined magnetic monopoles in a superconductor. We proceed to calculate the cost of this additional worldline suppression. 

The particle-like soliton is electrically charged under the massive vector field. This completes the duality of the superconductor; magnetic monopoles of the massive vector field are confined by `electrically' charged strings under the massive dual axion. The `magnetically' charged instantons of the dual axion are confined by electrically charged particles under the massive vector field. 

The confinement of instantons and their contributions to the path integral presents interesting model-building opportunities. A non-exhaustive list of models in which both the confined instantons and confining worldline are dynamical is provided in Section \ref{section3}. These include adaptations of models of particle confinement and the Green-Schwarz mechanism \cite{Green:1984sg}. We comment on the required spectrum of particles in such simple models, the connection to the massless up quark solution and the relevance for the strong CP problem.     

In other areas of the literature, axions with gauged shift symmetries have been discussed in  field theoretic contexts as dark matter candidates  \cite{Coriano:2018uip,Coriano:2010py}, at the LHC \cite{Irges:2007yha,Coriano:2009zh,Coriano:2010nf,Coriano:2008pg,Coriano:2005own,Dror:2017nsg,Dror:2018wfl,Dror:2017ehi,Dror:2017nsg}, as mixing with ordinary axions \cite{Shiu:2015uva,Shiu:2015xda,Choi:2019ahy,Berg:2009tg,Fraser:2019ojt}, in brane-bulk scenarios \cite{Cheng:2001ys} or as solutions to the strong CP problem \cite{Aldazabal:2002py}.

\setcounter{secnumdepth}{3}

\section{Massive dual axion}
\label{section1}
We begin by briefly reviewing the axion, the dual axion and the massive dual axion theory. Special importance is placed on the effect of this dual axion mass on the symmetries of the massless theory. Studying the fate of these symmetries will illuminate the fate of the charged objects of the theory, axion strings and instantons, in the superconducting phase. 

\subsection{Massless axion}
A massless axion is described by a compact scalar $a \equiv a + 2\pi$ with a Lagrangian of the form,
\begin{equation}
\mathcal{L} = \frac{F_{a}^{2}}{2}  \left(\partial_{\mu} a\right)^{2} \,.
\label{masslessaxion}
\end{equation}
Here $F_{a}$ is the fundamental period of the axion. This axion can be realized from the spontaneous breaking of a $U(1)^{(0)}$ symmetry\footnote{The notation used here $G^{(p)}$ is a standard convention adopted from the generalized symmetries literature for a symmetry group $G$ acting on gauge invariant objects of space-time dimensionality $p$. For our purposes, $p$ will always align with the number of indices on the associated gauge field that transforms under the shift.}, in which case the $a \equiv a+2\pi$ identification is an emergent gauge symmetry below the spontaneous breaking scale. 

The Lagrangian \eqref{masslessaxion} has a well-known non-linear $U(1)^{(0)}$ shift symmetry under which the axion transforms as
\begin{equation}
    a \rightarrow a + c^{(0)}, \quad \partial_{\mu}c^{(0)} = 0 \,.
\end{equation}
This shift symmetry is generated by the associated current
\begin{equation}
    j^{(0)}_{\mu} =   F_{a}^{2}  \partial_{\mu} a 
    \label{shiftsym}
\end{equation}
and is conserved by the equations of motion of \eqref{masslessaxion}. 

Instantons and other space-time configurations that are sources of a non-zero divergence of $j^{(0)}_{\mu}$ are said to be electrically charged under the axion. This follows identical conventions for the photon and the divergence of its field strength. The effect of such a source on the theory can be probed by an $e^{ia}$ (Wilson point) insertion in the path integral. The symmetry is explicitly broken if such sources have a finite creation action (are dynamical), such as gauge instantons.   

In four space-time dimensions, the Lagrangian \eqref{masslessaxion} also has an additional $2$-form $U(1)^{(2)}$ symmetry \cite{Gaiotto:2014kfa,Brennan:2020ehu,Choi:2022fgx,Heidenreich:2021xpr} associated with the conserved current 
\begin{equation}
j^{(2)}_{\mu \nu \rho} = \frac{1}{2 \pi} \epsilon_{\mu \nu \rho \lambda} \partial^{\lambda}a \,.
\label{twoformcurrent}
\end{equation}
The charge associated with the current \eqref{twoformcurrent} measures the winding number (monodromy) of the axion field around charged strings. In the frame of the axion, the conservation of this current is a topological condition ($dda = 0$) \cite{Heidenreich:2023pbi}. This should be contrasted with the conservation \eqref{shiftsym} of the axion shift symmetry by the equations of motion. This $U(1)^{(2)}$ symmetry will be associated with the shift symmetry of the dual $2$-form of the axion and will therefore be referred to as the dual axion shift symmetry.  

Axion strings and other space-time configurations that are sources of a non-zero divergence of $j_{\mu \nu \rho}^{(2)}$ are said to be magnetically charged under the axion. The insertion of a probe (non-dynamical) charged string worldsheet can be performed in the frame of the axion by excising a tube around the worldsheet and demanding a non-zero winding number on any curve linking this worldvolume. This requires the introduction of an unphysical Dirac sheet attached to the string. The $U(1)^{(2)}$ symmetry can be broken by the existence of dynamical strings with a finite creation action around which $d da \neq 0$.  

The axion shift symmetry $U(1)^{(0)}$ and dual axion shift symmetry $U(1)^{(2)}$ share a mixed 't Hooft anomaly \cite{Brennan:2020ehu}. Gauging one of the symmetries explicitly breaks the other. This mixed anomaly is key to understanding the often inverted behaviour of instantons and axion strings under deformations of the massless axion theory. If one of these symmetries is a gauge symmetry, sources that would break this symmetry are confined into charge neutral dipole configurations. The mixed anomaly implies that the other symmetry is explicitly broken. Sources that break that symmetry source a field profile that is screened at large distances. If the $U(1)^{(0)}$ symmetry is a gauge symmetry, this corresponds to the confinement of instantons and screening of axion strings. If the $U(1)^{(2)}$ symmetry is a gauge symmetry, axion strings are confined and the axion profile around instantons is screened. 

\subsection{Dual axion}
In four space-time dimensions, the theory \eqref{masslessaxion} of a massless compact scalar is dual \cite{Svrcek:2006yi} to that of a massless anti-symmetric two-index tensor field $B_{\mu \nu}$ with a Lagrangian
\begin{equation}
\mathcal{L} = \frac{1}{12f_{a}^{2}} H^{\mu \nu \rho}H_{\mu \nu \rho} , \quad H_{\mu \nu \rho} = \frac{1}{2}\partial_{[\mu} B_{\nu \rho]} 
\label{masslessdual}
\end{equation}
and a gauge identification
\begin{equation}
    B_{\mu \nu} \equiv B_{\mu \nu} + \partial_{[\mu} \Lambda_{\nu]} \,,
    \label{gaugeidentfy}
\end{equation}
for an arbitrary massless $U(1)$ gauge field $\Lambda_{\nu}$.

The normalization of the fundamental period of the dual $f_{a} = 2 \pi F_{a}$ is chosen \cite{Reece:2018zvv} such that the worldsheet $\Sigma$ of axion strings with unit winding number couples to $B$ as $\int_{\Sigma} B$. The inversion of the coupling strength $F_{a} \leftrightarrow \frac{1}{f_{a}}$ is sometimes called a `weak/strong' duality and is a general feature of dualizations \cite{Burgess:2023ifd}. 

A massless $2$-form with gauge identification \eqref{gaugeidentfy} has $1$ d.o.f. matching that of the original massless axion. The Lagrangian \eqref{masslessdual} also has both a $U(1)^{(0)}$ and $2$-form $U(1)^{(2)}$ symmetry. Under the $U(1)^{(2)}$ shift symmetry, the dual axion $B$ transforms as
\begin{equation}
    B_{\mu \nu} \rightarrow B_{\mu \nu} + c^{(2)}_{\mu \nu}, \quad \partial_{[\mu} c^{(2)}_{\nu \rho]} = 0 \,.
    \label{shiftsymmB}
\end{equation}
In this dual frame, the generating current 
\begin{equation}
    j^{(2)}_{\mu \nu \rho} = \frac{1}{2 f_{a}^{2}}\partial_{[\mu} B_{\nu \rho]} \,
    \label{dualtwosym}
\end{equation}
and is conserved by the equations of motion. This should be contrasted with 
the $U(1)^{(0)}$ symmetry in this frame, which is a
topological condition and generated by the current
\begin{equation}
    j^{(0)}_{\mu} = \frac{1}{4\pi}\epsilon_{\mu \nu \rho \lambda} \partial^{\nu} B^{\rho \lambda} \,.
    \label{dualzerosym}
\end{equation}
Configurations that are a source of non-zero $U(1)^{(2)}$ current are electrically charged under the dual axion. The effect of such  sources on the theory can be probed by an $e^{i \int B}$ (Wilson sheet) insertion along their worldsheet. Configurations that are a source of non-zero $U(1)^{(0)}$ current are magnetically charged under the dual axion. Non-dynamical configurations with $U(1)^{(0)}$ charge can be created in this frame by excising a small volume around a space-time point and demanding a non-zero $U(1)^{(0)}$ charge measured by any three sphere linking this point. This requires the introduction of an unphysical Dirac point (line in space-time).   

\subsection{Massive dual axion}
The aim of this section is to study the effect of mass contributions to the dual axion $B$. To this end, we add a mass $m$ to Lagrangian \eqref{dualtheory} through a generalization of the Proca/Stückelberg mechanism by adding a one-form $U(1)$ vector field $\Tilde{A}$ as,
\begin{equation}
\mathcal{L} = \frac{1}{12 f_{a}^{2}} H_{\mu \nu \rho} H^{\mu \nu \rho} - \frac{m^{2}}{4 f_{a}^{2}}\left(B_{\mu \nu}-\frac{f_{a}}{m} \partial_{[\mu}\widetilde{A}_{\nu]}\right)^{2} \,.
\label{massivetheory}
\end{equation}
Massive $2$-form fields with gauge identification \eqref{gaugeidentfy} have 3 d.o.f. and their properties have been studied in for instance \cite{
Cecotti:1987qr,Kuzenko:2020zad,Townsend:1981nu,Hell:2021wzm,Smailagic:2001ch} and in the context of cosmology in \cite{Capanelli:2023uwv}.

Several mass generating mechanisms which lead to a low-energy Lagrangian \eqref{massivetheory} exist for massless two-forms in field/string theories. Any strings coupled to the dual axion with either a finite creation action or which can be wrapped around compact $2$-cycles of the manifold give a mass to (components of) the dual axion. Such cycles can be absent in the IR, but the topology of spacetime seen by the dual axion can change in the UV, both in extra-dimensional theories and $4$D theories. A four dimensional mass for the dual axion can also be generated by the dimensional reduction of higher dimensional theories with couplings of the form $B^{2} F^{N}$ where $F$ is the field-strength of some $p$-form field. Activating fluxes $\int F = n$ on the internal/compactified cycles of the manifold will generate a mass $m \sim n^{N}$ \cite{Capanelli:2023uwv}. Such fluxes are standard in KKLT \cite{Kachru:2003aw} or Large Volume Scenarios \cite{Balasubramanian:2005zx} in string theory. Further examples of massive dual axions are those that participate in anomaly cancellation mechanisms such as the Green-Schwarz mechanism \cite{Green:1984sg}. In string theory, massive dual axions occur for instance in compactifications of heterotic string theory \cite{Svrcek:2006yi} or type II intersecting brane models \cite{Ibanez:2001nd,Blumenhagen:2006ci,Antoniadis:2002qm}. 

For completeness, we briefly mention that the massive dual axion theory can be thought of as the far IR realisation of a gauged $1$-form $U(1)^{(1)}$ shift symmetry of $\Tilde{A}$ that is spontaneously broken by the formation of a string condensate. This should be contrasted with theories in which the axion is massive and the path integral receives contributions from an instanton liquid instead. 

\subsection{Massive vector field}
Strings are `electrically' coupled to the dual axion and understanding their phenomenology in a phase in which the dual axion has a mass can be straightforwardly done from the massive dual axion perspective \eqref{massivetheory}, as will be done shortly in Section \ref{section2}. 

Instantons are `magnetically' coupled to the dual axion and understanding their fate is inevitably tied to understanding the fate of the original massless axion. The Lagrangian \eqref{massivetheory} is a dual description of several well-known models, one of which is a massive vector field, which has `eaten' the axion. We briefly review these dualization steps here. 

We first proceed by dualizing the vector field $\widetilde{A} \rightarrow A$, which yields the BF-theory,
\begin{equation}
    \mathcal{L} = \frac{1}{12 f_{a}^{2}} H_{\mu \nu \rho} H^{\mu \nu \rho} - \frac{1}{4e^{2}} F_{\mu \nu}F^{\mu \nu} + \frac{1}{4\pi} B_{\mu \nu} \widetilde{F}^{\mu \nu} \,.
    \label{bf}
\end{equation}
The coupling, field strength and dual field strength are normalized as
\begin{equation}
    e = \frac{2 \pi m}{f_{a}} \,, \quad F_{\mu \nu} = \partial_{[\mu} A_{\nu]} \,, \quad \widetilde{F}_{\mu \nu} = \frac{1}{2} \epsilon_{\mu \nu \alpha \beta} F^{\alpha \beta} \,.
\end{equation}
The BF coupling $B_{\mu \nu} \widetilde{F}^{\mu \nu}$ between the dual axion and vector field is a topological term in contrast to the kinetic coupling in \eqref{massivetheory} involving the dual vector field. This is another manifestation of duality; dynamic/kinetic couplings in one frame become topological in a dual frame \eqref{bf}. 

The topological coupling in \eqref{bf} implies that the axion is eaten by the vector field $A$. This can be seen by also dualizing the dual axion $B \rightarrow a$, which yields the Proca/Stückelberg action of a massive vector field,
\begin{equation}
\mathcal{L} = - \frac{1}{4 e^{2}} F_{\mu \nu}F^{\mu \nu} +  \frac{F_{a}^{2}}{2}\left( \partial_{\mu}a -  A_{\mu} \right)^{2}  \,. 
\label{dualtheory}
\end{equation}
The original axion becomes embedded as the longitudonal mode of a $U(1)$ gauge field $A$ matching the 3 d.o.f. of theory \eqref{massivetheory}. The vacuum acts as a superconductor and the fate of instantons in this phase is discussed in Section \ref{section3}.

\section{Axion strings}
\label{section2}
We proceed by studying the fate of axion strings and instantons in the superconductor, starting with axion strings. 

Strings can be electrically coupled to the dual axion $B_{\mu \nu}$. If such a string is described by a current density $J^{\mathrm{string}}_{\mu \nu}$, then this coupling is facilitated by the interaction 
\begin{equation}
    \mathcal{L} \supset \frac{1}{2}B^{\mu \nu} J^{\mathrm{string}}_{\mu \nu} \,.
\end{equation}
Such strings act as sources of field strength $H_{\mu \nu \rho}$ `electric flux' lines. 

In the presence of a mass for the dual axion, these flux lines can end in the string condensate and become screened. This can be observed by studying 
the equations of motion of \eqref{massivetheory} in the gauge $d\Tilde{A} = 0$, which become
\begin{equation}
\label{inverting}
    \partial^{\rho} H_{\rho \mu \nu} =  -m^{2} B_{\mu \nu} + f_{a}^{2}J^{\mathrm{string}}_{\mu \nu}   \,.
\end{equation}
The mass $m$ suppresses any static long-distance non-zero $B_{\mu \nu}$ fields and far-separated strings become non-interacting.

This can be confirmed by studying a string that has unit charge and is oriented in the $z$-direction of a cylindrical coordinate system $(t,z,r,\phi)$ such that $J^{\mathrm{string}}_{tz} = \delta(r) $. We can solve for the profile around a charged string by inverting \eqref{inverting} as
\begin{equation}
B_{\mu \nu} = \frac{f_{a}^{2}}{\partial^{2}+ m^{2}} \left( J^{\mathrm{string}}_{\mu \nu} - \frac{1}{m^{2}} \partial^{\rho} \partial_{[\mu}  J^{\mathrm{string}}_{\nu] \rho} \right) \,.
\end{equation}
Solving around such a profile yields a radially decreasing dual axion field
\begin{equation}
    B_{tz} = -\frac{f_{a}^{2}}{2\pi} K_{0}(mr)  \,.
    \label{axionstrings}
\end{equation}
Here $K_{i}(x)$ is the i'th modified Bessel function. For distances $r \gg \frac{1}{m}$, the dual axion field is screened exponentially.  

The monodromy around axion strings is no longer conserved as the $U(1)^{(2)}$ shift symmetry \eqref{shiftsymmB} of the dual axion is explicitly broken by the mass. The winding number of a string at a given radius $r$ is given by integrating \eqref{dualtwosym} around the string, which for profile \eqref{axionstrings} implies
\begin{equation}
\label{windingmode}
Q_{\mathrm{wind}} =  mr K_{1}(mr)  \,.
\end{equation}
Close to the string, the winding number is constant. For distances $r \gg \frac{1}{m}$, the dual axion field is screened and the winding number is exponentially small. 

In the dual Proca theory \eqref{dualtheory} of a superconductor, the screening of the dual axion field profile is understood as follows. The axion profile winds around the string $da \neq 0$. Far away from the string core, the axion profile is compensated by a gauge field such that the axion kinetic term $ \frac{F_{a}^{2}}{2}\left( \partial_{\mu}a - A_{\mu} \right)^{2}$ vanishes. The tension of the string due to the winding of the axion only receives contributions from the string core and therefore far-separated strings become non-interacting. 

The phenomenology of such strings is that of gauge strings and the screening of the dual axion profile does not imply that axion strings in a superconductor are unobservable. As the kinetic term vanishes, the magnetic flux carried by the string is non-zero: $\int A = \int da$. This magnetic flux results in an Aharanov-Bohm phase for any charged particle winding around the string. 

Further interactions of gauge strings with their environment can come from coupling to other long-range fields such as electromagnetism or gravity. 

One of the earliest phenomenological signatures of such gauge strings focused on their gravitational signal, which is set by the string's tension $T \sim F_{a}^{2}$. Oscillating loops of such strings produce gravitational waves \cite{PhysRevD.31.3052,PhysRevD.42.354} with the total power $P$ being radiated by oscillating loops of strings given by $P \sim G F_{a}^{4}$ where $G$ is Newton's constant. Additionally, such strings can form seeds for cosmological perturbations  \cite{Kibble:1980mv,Zeldovich:1980gh,Vilenkin:1981iu}. As a string of length $R$ has a mass $R F_{a}^{2}$, the gravitational effects of such strings are only observable for very large scales $F_{a}$ \cite{Witten:1984eb}.

A secondary, non-gravitational interaction of the gauge string with the environment can come from the physics at the core of the string. This requires a description of the physics at the core of the string and a distinction has to be made between a string formed in an abelian Higgs mechanism and a so-called fundamental or Stückelberg string \cite{Reece:2018zvv}. Both admit Lagrangians \eqref{massivetheory} far away from the core, but only the former string has a core at which the shift symmetry is restored.

Zero modes can reside at the cores of both strings. These zero modes can carry electric charge, making the string superconducting \cite{Witten:1984eb}. If such a string passes through galactic magnetic fields, large currents can be induced on the string. These currents in turn create a large magnetic field, which could be observed through radiation or through the synchotron radiation of trapped charged particles. 

Stückelberg strings are still accompanied by both an axion and a radial mode, but the vanishing point of this radial mode lies at an infinite distance in field space. There is no symmetry restoration at the core, but instead local effective field theory is expected to break down near the core of the string \cite{Reece:2018zvv,March-Russell:2021zfq} and their phenomenology might be vastly different.

\section{Instantons}
\label{section3}
Instantons can be electrically coupled to the axion $a$. In theories in which the axion shift symmetry is gauged, instantons no longer couple to just axion insertions but instead couple to the gauge-invariant operator
\begin{equation}
    \mathrm{Exp}\left[i a - i  \int_{C} A\right] \,.
    \label{wilsonline}
\end{equation}
where $C$ is any curve extending from the instanton to infinity and $e = \frac{m}{F_{a}}$. The presence of this additional Wilson line specifies the worldline of a charged (non-dynamical) particle stretching from the instanton. The space-time process describes the creation/decay of the particle at the instanton. The additional worldline cost of this charged particle to the partition function $Z$ of any isolated instanton confines this instanton. Only gauge invariant combinations in which the worldline $C$ stretches between different instantons contribute to the vacuum to vacuum partition function. Instantons will still contribute to amplitudes in which the Wilson lines can end on the boundary, such as those that involve the confining particles in the initial or final states. 

\begin{figure}
\begin{tikzpicture}
\begin{scope}[shift={(0,0)},scale=1.2, transform shape]
    \draw[->] (0,0) -- (1.75,0);
    \draw[->] (0,0) -- (0,1.5);
     \draw[color=orange, thick, line cap= rect, line join= bevel,
rounded corners=8pt] (0.5,0.5) .. controls (0.8,0.8) and (1.2,0.1) .. (1.5,0.5);
     \draw[color=orange,rotate around={45:(0.5,0.5)}, thick, line cap= rect, line join= bevel,
rounded corners=8pt] (0.5,0.5) .. controls (0.8,0.8) and (1.2,0.1) .. (1.5,0.5);
 \draw[color=orange,rotate around={20:(0.5,0.5)}, thick, line cap= rect, line join= bevel,
rounded corners=8pt] (0.5,0.5) .. controls (0.8,0.8) and (1.2,0.1) .. (1.5,0.5);
     \filldraw[red] (0.5,0.5) circle (2pt) node[anchor=east] (p1) {I};
         \path (1.25,-0.25) node (p3) {space};
    \path (-0.5,1) node (p4) {time};
    \end{scope}
    \begin{scope}[shift={(4,0)},scale=1.2, transform shape]
    \draw[->] (0,0) -- (1.75,0);
    \draw[->] (0,0) -- (0,1.5);

     \draw[color=orange,rotate around={0:(0.5,0.5)}, thick, line cap= rect, line join= bevel,
rounded corners=8pt] (0.5,0.5) .. controls (0.8,0.9) and (1.2,0.4) .. (1.5,1);
    \filldraw[darkblue] (1.5,1) circle (2pt) node[anchor=west] (p2) {AI};
     \filldraw[red] (0.5,0.5) circle (2pt) node[anchor=east] (p1) {I};
    
    \path (1.25,-0.25) node (p3) {space};
    \path (-0.5,1) node (p4) {time};
    \end{scope}
\end{tikzpicture}
\caption{\textit{Left}: Space-time diagram of an instanton (I) (red) confined by particle-like soliton worldlines (orange) describing the creation/decay of such particles at the instanton vertex. \textit{Right}: Space-time diagram of an instanton (I) (red) and anti-instanton (AI) (blue) configuration with a particle-like soliton worldline (orange) stretched between the instantons describing the creation and subsequent decay of the particle.}\label{fig1}
\end{figure}
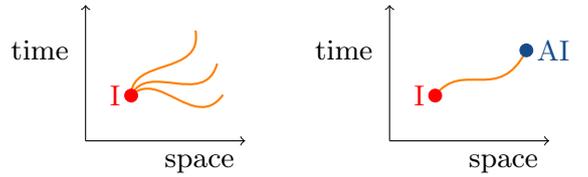

This confinement of instantons is analogous to the confinement of particles by Abrikosov/Nielsen-Oleson vortex solutions \cite{Abrikosov:1956sx,Nielsen:1973cs} stretching between magnetic monopoles in a superconductor or between electric charges in models with a massive dual photon \cite{Hook:2022pcf}. An artistic impression for the case of a single instanton and an instanton-anti-instanton dipole is provided in figure \ref{fig1}.

Confinement of instantons is not unique to \eqref{massivetheory} and has a long history. One of the earliest examples of instanton confinement-deconfinement transitions occurs in $2$D (space-time dimensions) in the well-known $XY$ model. This model has the coarse-grained description of a simple massless compact scalar. Vortices or strings of this compact scalar are instantons in $2$D and are confined by a logarithmic potential, forming bound vortex pairs. Increasing the temperature allows for a transition to a gas of free unbound vortices above a critical temperature described by the infinite order Berezinskii–Kosterlitz–Thouless \cite{Berezinsky:1970fr,Kosterlitz:1974nba} phase transition.

Confinement of instantons in the presence of massless fermions also played an important role in the early studies of the vacuum structure of QCD \cite{Callan:1976je,Callan:1977gz}. In his seminal paper, 't Hooft showed \cite{tHooft:1976snw} that the existence of normalizable zero mode solutions to the massless Dirac equation implied the vanishing of the single instanton partition function in the presence of massless fermions. The exchange of massless fermions gives rise to strong logarithmic interactions between instantons and anti-instantons, leading to logarithmic confinement \cite{Callan:1977gz,Lee:1979sm}. Instanton and anti-instantons form closely bound dipole pairs and vacuum tunneling effects become negligible in the absence of chiral symmetry-breaking sources. Instantons would instead contribute to correlation functions that include the fermion zero mode, such as a chiral condensate. 

Confinement of instantons is crucial for extended particle-like field configurations that enjoy topological protection, like vortices and monopoles. Smooth deformations of their field profile in time cannot change the topological index associated with these solitons, and their number is conserved. However, singular field configurations can describe quantum tunneling events between these distinct topological sectors. Therefore, an additional prerequisite to the number conservation of such solitons is the confinement of the associated instantons. This is most pronounced for topological defects that lack core singularities, such as Skyrmions. For instance, unconfined instantons exist in $3$D in the $\mathbb{C}P^{1}$ model describing a $S^{2}$ valued field $\phi$ on a compact space $S_{2}$ \cite{Polyakov:1975yp}, where instantons describe the sudden decay of the Skymrion and violation of Skyrmion number. However, not all such defects are confined. Hopfions \cite{Faddeev:1975tz,Faddeev:1996zj}, non-trivial maps $\pi_{3}(S^{2})$ from a compact space $S^{3}$ to a $S^{2}$ valued field do have an associated conserved charge in $4$D space-time. This is due to the logarithmic confinement of the associated Hopf instantons. A general discussion on the stability of topological defects due to instanton confinement and the renormalization flow of instanton interactions can be found in the recent paper \cite{Nikolic:2023uxw}.

Let us briefly comment that this instanton suppression cannot solve the axion quality problem of a secondary massless axion. In the presence of multiple species of instantons, a diagonal combination of instantons is uncoupled to the axion with a dual mass. This diagonal combination will remain unsuppressed and can give a mass to a secondary axion.  
  
\subsection{Instanton gas}
\label{twoinstantons}
Confinement forces instantons into gauge-invariant dipole configurations such as the instanton-anti-instanton dipole depicted in figure \ref{fig1}. Analogous to flux tubes stretching between confined particles producing a linear potential, the existence of the worldline of the charged particle implies an additional cost to the instanton-anti-instanton partition function. We proceed by calculating this additional cost. 

In a system consisting of a gas of instanton-anti-instanton configurations, in which the worldline would be dynamical, this worldline would relax to the shortest distance between any instanton-anti-instanton pair. In the massive dual axion theory \eqref{massivetheory}, both the worldline and the instantons are non-dynamical probe objects. We will rectify this in the next subsection by introducing additional degrees of freedom to \eqref{massivetheory} in order to make both the instantons and worldlines dynamical, but for now we wish to describe only this steady state situation in which the worldline has relaxed to the shortest distances between the instantons.

There is an equivalent formulation of the coupling \eqref{wilsonline} first introduced by Zwanziger \cite{Zwanziger:1970hk} for the analogous electromagnetic coupling of photons to magnetic monopoles, which will be useful to describe this instanton-anti-instanton dipole. In this formulation, one introduces axion shift symmetry sources by an instanton density $J_{\mathrm{inst}}$. To each source, one attaches a Dirac line described by a unit vector $n^{\mu}$. The coupling between the density and the axion is then achieved by the appropriate addition of the interaction Lagrangian
\begin{equation}
\mathcal{L}_{\mathrm{int}} =   \left(a - \frac{n^{\mu}}{n \cdot \partial} A_{\mu} \right) J_{\mathrm{inst}} \,.
\label{couplingnew}
\end{equation}
In the limit $m \rightarrow 0$ keeping $F_{a}$ fixed, the coupling $e$ of the photon vanishes and one returns to a theory of a compact scalar $a$, the axion, coupled to instantons. When $m \neq0$, the Dirac line becomes physical and is identified with the worldline of the non-dynamical particle. Since the worldline of the particle is non-dynamical, the interaction \eqref{couplingnew} is Lorentz violating.

In the gauge $a = 0$, the system 
in the Zwanziger formulation \eqref{couplingnew} is described by an effective non-conserved current
\begin{equation}
    J^{\mu} = \frac{n^{\mu}}{n \cdot \partial} J_{\mathrm{inst}} \,,
    \label{effectivecurrent}
\end{equation}
coupled to massive vector field $A_{\mu}$ with corresponding equations of motion in Euclidean space-time
\begin{equation}
 A^{\mu} = \frac{g^{\mu \nu} - \frac{\partial_{\mu} \partial_{\nu}}{m^{2}}}{-\partial^{2} + m^{2}}  e^{2}J^{\nu} \,.
\end{equation}
The current \eqref{effectivecurrent} describes the worldline of a charged particle in the direction $n^{\mu}$ that emanates from a non-zero instanton density $J_{\mathrm{inst}}$. This should be compared to the equivalent Wilson line formulation \eqref{wilsonline} which requires the introduction of a non-conserved current $J^{\mu}$ along the curve $C$ coupled to the massive gauge field $A^{\mu}$ that emerges from an insertion of $e^{ia}$. 

We are interested in the field configuration in Euclidean space-time describing instanton-anti-instanton dipole configuration with a separation length $R$ in time. In a spherical coordinate system $(t,r,\phi,\theta)$, such a configuration is described by the instanton density $J_{\mathrm{inst}} = \delta(x^{\mu}+\frac{R}{2} \hat{t}) - \delta(x^{\mu}-\frac{R}{2}\hat{t})$ and corresponds to an effective current \eqref{effectivecurrent} of the form
\begin{equation}
J^{0} =  \delta(r) \mathrm{Rect}\left(\frac{t}{R}\right)
\label{explicitcurrent}
\end{equation}
in which the Dirac line $n = \hat{t}$ stretches between the instantons and $\mathrm{Rect}\left(\frac{t}{R}\right)$ is a unit box with size $R$ centered at $t=0$. This current describes the existence of a charged particle that sits at the origin for a time $R$.

In the limit $R \rightarrow \infty$, the profile for $J^{\mu}$ is that of a charged particle sitting at the origin for all time and one has the ordinary screened solution (in the normalization of \eqref{dualtheory}) for a charged particle,
\begin{equation}
A^{0} = \frac{e^{2}}{4 \pi r}e^{-mr} \,.
\end{equation}
At a finite distance $R$, we can calculate the contribution to the partition function $Z$ due to such an instanton-anti-instanton configuration. Given an effective current $J^{\mu}$, such an evaluation of the partition function amounts to completing the square of \eqref{dualtheory} as,
\begin{equation}
\ln{Z} = -\frac{e^{2}}{2} J_{\mu}\left(\frac{   g^{\mu \nu} - \frac{\partial^{\mu} \partial^{\nu}}{m^{2}}}{-\partial^{2} + m^{2}} \right)  J_{\nu} \,.
\end{equation}
In the Wilson line formulation \eqref{wilsonline} and the gauge $a=0$, this simply amounts to calculating the expectation value of $e^{i \int_{C} A}$ along the contour $C$ stretched between the instantons. 

We proceed by plugging in the explicit expression \eqref{explicitcurrent} for $J^{\mu}$, 
\begin{equation}
    \ln{Z} = - \frac{e^{2}}{2} \int_{-\frac{R}{2}}^{\frac{R}{2}} dt \int_{-\frac{R}{2}}^{\frac{R}{2}} dt' \ G(t-t') \,.
\end{equation}
Here, $G(t)$ is the Green's function of the massive vector field. Plugging in the expression for this Green's function in momentum space, introducing the three momentum $\Vec{k}$ as $k_{\mu} = (k_{0},\Vec{k})$ and performing the temporal integrals gives
\begin{equation}
    \ln{Z} = - 2e^{2} \int \frac{dk^{0}}{2\pi} \frac{d^{3} \Vec{k}}{(2\pi)^{3}}  \frac{\sin^{2}{\left(\frac{k_{0}R}{2}\right)} }{k_{0}^{2}}\frac{1 - \frac{k_{0}^{2}}{m^{2}}}{k^{2} + m^{2}} \,.
\end{equation}
Performing the temporal momentum integrals, keeping only the $R$-dependent pieces and splitting the partition function for later convenience yields
\begin{equation}
    \label{partitionfunction}
    \ln{Z} =  - \frac{e^{2}}{2}S - \frac{e^{2}}{(2 \pi)^{2}} \frac{K_{1}(mR)}{mR} \,,
\end{equation}
where $S$ is dominated by the action of the charged soliton,
\begin{equation}
\label{selfenergy}
    S =  \int \frac{d^{3} \Vec{k}}{(2\pi)^{3}} \left(\frac{R}{\Vec{k}^{2} + m^{2}} + \frac{e^{- R \sqrt{\Vec{k}^{2} + m^{2}}}}{\left(\Vec{k}^{2} + m^{2} \right)^{\frac{3}{2}}}   \right)\sim R \Lambda \,.
\end{equation}
From the partition function \eqref{partitionfunction} it is clear what the interpretation of the addition of the current $J_{\mathrm{inst}}$ to the system is. In the limit $mR \ll  1$ keeping $F_{a}$ fixed, the second term in \eqref{partitionfunction} dominates $\ln{Z} \sim -\frac{1}{F_{a}^{2}R^{2}}$. This simply describes a massless axion propagating between the instanton and anti-instanton. 

 In the opposite limit $m R \gg 1$ keeping $F_{a}$ fixed, the massive photon is coupled to the current $\frac{n^{\mu}}{(n \cdot \partial)} J_{\mathrm{inst}}$. This current describes a unit box or equivalently, the insertion of a charged particle for a time $R$. The partition function of such a configuration scales as $\ln{Z} \sim - e^{2}R \Lambda$ where the UV-cutoff $\Lambda$ is the electric self-energy of the charged particle. The instantons are linearly confined. 

Similar to the confinement of colour, the presence of instantons with a finite action $e^{-S_{0}}$, reduces the effective range of the confining contribution. In the presence of such instantons, the worldline of the charged particle can be broken by a dipole instanton-anti-instanton pair. Such a breaking becomes favourable whenever the distance $e^{2} R \gtrsim \frac{S_{0}}{\Lambda}$ and therefore the confining contribution is only of finite range. This length is further reduced by virtual instanton fluctuations \cite{Nikolic:2023uxw}. 

We briefly mention that these results could have equivalently been obtained from a dimensional reduction of ordinary particle confinement. Loops of particles along the extra dimension are instantons in the reduced theory. The partition function $\ln Z$ in $4$D of these instantons is the potential $V_{5}$ in the $5$D theory of the looping particle. If the particles are charged under a gauge field in $5$D, this potential can either be Coulomb $V_{5}(R) \sim \frac{1}{R^{2}}$ for a massless photon or confining  $V_{5}(R) \sim R$ for a massive dual photon \cite{Hook:2022pcf}.

\subsection{Confinement of instantons}
\label{instantonconfinement}
Both the instanton and confining worldline can be made dynamical by introducing additional degrees of freedom to the massive theory \eqref{massivetheory}. These confining degrees of freedom come in two flavours, bosons and fermions. 

\subsubsection{Bosonic confinement}
Simple models of confinement of instantons by worldlines of bosons can be obtained in $3$D. We will focus on the confinement of magnetic monopoles, which are instantons in this dimension and are associated with a non-trivial second homotopy index $\pi_{2}$. As $\pi_{2}(G)=1$ for any compact, connected Lie group $G$ \cite{Cartan}, such instantons only form when a gauge group is spontaneously broken $G \rightarrow H$ and are associated with the index $\pi_{2}(G/H)$. Subsequent breaking of $H$ will always eventually result in confinement of such instantons as they should be absent from the full breaking $\pi_{2}(G/1) = 1$. 

Consider simple two-step breaking models in which a gauge group $G$ breaks as $G \rightarrow H_{1} \rightarrow H_{2}$. We will focus on models in which instantons that form in the first breaking step confine due to meta-stable particles that form in the second breaking step $H_{1} \rightarrow H_{2}$. Particles formed in the second breaking step are classified by their topological index $\pi_{1}(H_{1}/H_{2})$. In the larger gauge group $G$, such configurations are described by the topological index $\pi_{1}(G/H_{2})$ and for each particle there exists a mapping
\begin{equation}
    \pi_{1}(H_{1}/H_{2}) \rightarrow \pi_{1}(G/H_{2})
\end{equation}
If this mapping has a non-trivial kernel, then
two distinct particles $p_{1}$ and $p_{2}$ associated with distinct non-trivial loops in $H_{1}/H_{2}$ can be continuously connected in $G/H_{2}$. In such cases, there exists an instanton associated with an element in $\pi_{2}(G/H_{1})$ describing a surface in $G/H_{2}$ that connects the two loops. The winding configurations are said to be able to unwind in the larger gauge group $G$ and the instanton allows for the particle transition/decay $p_{1} \rightarrow p_{2}$. If a particle is mapped to the identity in $\pi_{1}(G/H_{2})$, then the instanton describes the decay of the particle to the vacuum. A full account of the decays of metastable topological defects in any dimension can be found in \cite{Preskill:1992ck}. 

If the instanton is absent from the total breaking $\pi_{2}(G/H_{2}) = 1$, then the instanton only exists as an endpoint or kink or twist on the worldline of meta-stable particle(s). Simple examples of such models are breaking patterns
\begin{equation}
    G \rightarrow U(1) \rightarrow 1
    \label{simplebreaking}
\end{equation}
for compact, connected and simply connected Lie groups $G$. Particles form during the second breaking step $\pi_{1}(U(1)) = \mathbb{Z}$, but can unwind in the larger simply connected gauge group as $\pi_{1}(G) = 1$. The decays/transitions of these particles are facilitated by instantons. The instantons are absent from the full breaking and instead are confined as the end-points of such particle worldlines. Similar considerations apply to instantons confined on the worldlines of non-Abelian particles such as in \cite{Tong:2003pz,Shifman:2004dr}. 
 
A well-known example is magnetic monopoles (instantons in this dimension) that form in $SU(2) \rightarrow U(1) \rightarrow 1$ breaking patterns by subsequently one adjoint and one fundamental scalar. During the second breaking by the fundamental scalar, non-perturbative particles form $\pi_{1}(U(1)) = \mathbb{Z}$, which are time-independent configurations in which the fundamental scalar winds. In the full breaking pattern, all such winding configurations are trivial $\pi_{1}(SU(2))=1$ and can therefore be unwound by instanton insertions $\pi_{2}(SU(2)/U(1)) = \mathbb{Z}$. These instantons are 't Hooft-Polyakov monopoles \cite{tHooft:1974kcl,Polyakov:1974ek} that form during the first breaking of $SU(2) \rightarrow U(1)$ by the adjoint scalar. These instantons are absent in the full breaking $\pi_{2}(SU(2)) = 1$ and only occur at the end-points of the worldlines of the non-perturbative particles.    

If the second breaking in \eqref{simplebreaking} is instead to a discrete subgroup $Z_{N} \subset U(1)$, then the instantons become kinks or twists on the worldline of particles. A well-known example is magnetic monopoles, which are kinks on particle worldines in $SU(2) \rightarrow U(1) \rightarrow \mathbb{Z}_{2}$ double breaking models \cite{PhysRevLett.55.2398} by two orthogonal adjoints. During the second breaking $U(1) \rightarrow \mathbb{Z}_{2}$ by an adjoint $\phi$, particles form $\pi_{1}(U(1)/\mathbb{Z}_{2}) = \mathbb{Z}$  as winding configurations of $\phi$. In the full breaking, there are only two topologically inequivalent winding configurations: $\pi_{1}(SU(2)/\mathbb{Z}_{2}) = \mathbb{Z}_{2}$. The magnetic monopole formed during the first breaking connects the worldlines of these configurations. No closed loop can be formed and the instanton is confined.  

Both of these models provide examples of completions of the massive dual axion Lagrangian \eqref{massivetheory} in $3$D in which both the instanton and worldline are dynamical. We identify the axion with the magnetic dual of the $U(1)$ photon of these models. After the first breaking step, monopoles (instantons) form which are magnetically charged and couple to the axion/dual photon. A gas of these instantons would have given a mass $m\sim e^{-S_{\mathrm{mon}}}$ to the axion/dual photon \cite{Polyakov:1976fu}. 

However, this is prevented by the second breaking step at scales $v$, where particles associated with the winding of the second scalar $\phi$ form attached to the monopole. These particles come attached to any 't Hooft vertex associated with the monopole and the monopole becomes confined. The photon (dual axion) obtains a mass from eating the second scalar $\phi$. The partition function of any pair of instantons separated by a space-time distance $R$ is suppressed by the action cost of the charged particle travelling between the instantons. As the non-perturbative particle has mass $v$ (which includes the self-energy), this results in a partition function suppression of $\ln{Z} \sim -v R$. 

Extending such models of instanton confinement to $4$D, one runs into issues. In this dimension, we consider instantons associated with the topological index $\pi_{3}$ and again wish to explore the confinement of such instantons by particle worldlines. This time instantons already exist; $\pi_{3}(G) \neq 0$ for a generic non-Abelian gauge group $G$. In two-step breaking models, we need to follow both such instantons and possible additional instantons created in the breaking associated with the index $\pi_{3}(G/H_{1})$.

Associated with the first breaking $G \rightarrow H_{1}$, there is a fibration $H_{1} \rightarrow G \rightarrow G/H_{1}$ and a long exact sequence \cite{Csaki:1998vv} in homotopy
\begin{equation}
   \cdots \rightarrow \pi_{3}(H_{1}) \rightarrow \pi_{3}(G) \rightarrow \pi_{3}(G/H_{1}) \rightarrow \pi_{2}(H_{1}) = 1 \rightarrow \cdots 
\end{equation}
The exactness of the first three entries of the sequence implies that the instantons of $G$ either descend to an instanton in the unbroken gauge group $H_{1}$ or form a subset of the instantons associated with non-trivial elements of the index $\pi_{3}(G/H)$. In fact, the exactness of the last three entries and the vanishing index $\pi_{2}(H_{1}) = 1$ implies that no other instantons are created during the breaking process, 
\begin{equation}
\pi_{3}(G/H_{1}) \equiv \pi_{3}(G)/\pi_{3}(H_{1}) \,.
\end{equation}
All instantons descend from those of $G$ and split into instantons associated with either the unbroken part of the gauge group $\pi_{3}(H_{1})$ or the broken part $\pi_{3}(G/H_{1})$.

A simple example involves the breaking of $SU(2) \rightarrow U(1)$ by an adjoint scalar. The  fibration associated with the breaking is the Hopf fibration $S^{1} \rightarrow S^{3} \rightarrow S^{2}$. The low-energy gauge group $U(1)$ has no instantons, instead $\pi_{3}(SU(2)/U(1)) =\mathbb{Z} = \pi_{3}(SU(2))$. 

\textit{What happens to these instantons during the second breaking $H_{1} \rightarrow H_{2}$?} Particle-like defects can form during the second breaking, this time associated with the topological index $\pi_{2}(H_{1}/H_{2})$. In the larger gauge group $G$, the worldlines of particles that can end on instantons are associated with the kernel of the mapping
\begin{equation}
    \pi_{2}(H_{1}/H_{2}) \rightarrow \pi_{2}(G/H_{2}) \,.
\end{equation}
However, the kernel of such a mapping must always be trivial \cite{Preskill:1992ck} as $\pi_{2}(H_{1}) = 1$ for all compact finite-dimensional Lie groups $H_{1}$. Any magnetic monopole associated with $\pi_{2}(H_{1}/H_{2})$ remains absolutely stable when embedded into a larger gauge group $G$. If instantons associated with the index $\pi_{3}(G/H_{1})$ do connect particle worldlines, then these worldlines are topologically equivalent with respect to the index $\pi_{2}(H_{1}/H_{2})$.

Even-though the topological index of instantons is protected against confinement, there could still be dynamical reasons for the absence or confinement of instantons.  Upon higgsings, gauge instantons in $4$D in the broken part of the gauge group become constrained instantons \cite{Affleck:1980mp}. By a generalization of Derrick's theorem \cite{Derrick:1964ww}, such classical instanton solutions do not exist in the presence of a non-trivial scalar \cite{Manton:2004tk} in the absence of any scale-fixing source \cite{Affleck:1980mp}. Due to the Higgs phase of the bulk, any instanton solution with a non-trivial scalar profile wants to shrink to minimize its action. 

\textit{So what happens to instantons associated with $\pi_{3}(G/H_{1})$ during the first breaking?} Quantum corrections, such as asymptotic freedom, can prefer larger instanton solutions. A competition between these two effects can lead to constrained instantons  \cite{Affleck:1980mp}; approximate instanton solutions, to still dominate the functional integral \cite{tHooft:1976snw} and such instantons are unconfined. 

Instantons can also survive the Higgs phase of the bulk by residing as moduli degrees of freedom on the worldvolume of larger host solitons \cite{Eto:2004rz}. The action of these instantons diverges with the infinite extent of the host soliton, but there is no topological protection against compactifying the worldvolume. If the host soliton worldvolume is compact, any charge associated with the host vanishes, and only the instanton number remains. 

Examples of instantons residing on host solitons are worldlines of magnetic monopoles that carry a $U(1)$ modulus rotor degree of freedom. A twist or winding of this rotor degree of freedom on the magnetic monopole loop carries instanton number \cite{Jackiw:1975ep} through their Hopf invariant \cite{Bruckmann:2002jm,Jahn:1999wx}. Such examples include $SU(2)$ gauge theories spontaneously broken to $U(1)$, which admit 't Hooft-Polyakov monopoles as $\pi_{2}(SU(2)/U(1)) = \mathbb{Z}$ and instantons $\pi_{3}(SU(2)/U(1)) =\mathbb{Z}$. These instantons live on host magnetic monopole worldlines. Since the $U(1)$ modulus descends from an unbroken $U(1)$ gauge symmetry in the bulk, the modulus does not admit a potential and the instanton is diluted along the monopole worldline. The action for a loop of such a magnetic monopole with Schwinger length $\ell$ includes a cost $m_{M}^{2} \ell$ for a monopole with mass $m_{M}$, which is balanced against a gradient energy from the winding of the instanton $\sim \frac{1}{\ell}$ around the loop. Contributions of such instantons to the axion mass were recently calculated in \cite{Fan:2021ntg}.   

Additional examples of instantons on host solitons exist in $2$D where $Q$-lumps are twists on $Q$-kinks \cite{Abraham:1992qv}. In $4$D, $\mathcal{N}=2$ SQCD models in the Higgs phase \cite{Hanany:2004ea} have Yang-Mills instantons that reside as lumps on the worldsheet of non-Abelian vortex strings \cite{Tong:2003pz,Eto:2004rz} or as stabilized Skyrmions on domain wall worldsheets \cite{Eto:2005cc}. Further examples can be found in \cite{Nitta:2013vaa,PhysRevD.87.066008,Nitta:2022ahj} and references therein. We should mention that the exact relation (see for instance \cite{Brower:1996js}) between these hosted instantons and the constrained instantons of \cite{Affleck:1980mp} 
is still an open problem and deserves further study.

\subsubsection{Fermionic confinement}
\label{fermionconfinement}
Instantons can become confined by the existence of fermion zero modes. By the Atiyah-Singer index theorem \cite{Atiyah:1963zz}, the difference between the number of left and right chiral solutions to the massless Dirac equation on an even-dimensional compact space is given by the instanton number. In his seminal paper, 't Hooft showed \cite{tHooft:1976snw} that the existence of normalizable zero mode solutions to the massless Dirac equation in the background of an instanton implied the vanishing of the single instanton partition function. Instead, the instanton will contribute to correlation functions that include the fermion zero mode, such as the chiral condensate. 

Instanton-anti-instanton dipoles will receive contributions to their partition function from both the exchange of gluons and massless fermions. The gluonic exchange gives rise to dipole-dipole interactions $\ln{Z} \sim -\frac{1}{R^{4}}$ where $R$ is the distance between the instanton and anti-instanton \cite{Callan:1977gz,Schafer:1996wv}. The exchange of massless fermions gives rise to strong logarithmic interactions $\ln{Z} \sim - \ln{R}$ between the instanton and anti-instanton, leading to logarithmic confinement \cite{Callan:1977gz,Lee:1979sm}. 

Coupling such logarithmically confined instantons to a gauged axion results in the well-known Green-Schwarz mechanism \cite{Green:1984sg} and another realization of a superconductor in which both the confined instantons and confining particles are dynamical.

Consider coupling the massive dual axion or gauged axion to instantons. These instantons can be those of the massive vector field or possibly an external gauge group $G$, which we take to have traceless generators for simplicity. This can be done by a coupling\footnote{The allowed values of $c_{a}$ and $c_{G}$ depend on the spacetime manifold and the global gauge group, see \cite{Choi:2023pdp,Reece:2023iqn}. For simplicity, we pick a spin manifold and the product group $U(1)^{0} \times G$, which implies $(3c_{a},c_{G}) \in \mathbb{Z} \times \mathbb{Z}$}
\begin{equation}
   \mathcal{L} \supset c_{a} \frac{a}{16\pi^{2}} F \widetilde{F} +  c_{G}\frac{a}{16 \pi^{2}} \mathrm{tr} G \widetilde{G} \,.
   \label{axionshiftanomaly}
\end{equation}
This coupling is anomalous under the gauged  $U(1)^{0}$ shift symmetry $a \rightarrow a + \lambda$. In order to cancel the anomaly, one introduces a set of massless Weyl fermions charged under both the non-Abelian gauge group $G$ and the gauged axion shift symmetry $U(1)^{0}$. One can choose the anomalous variation of the fermion measure to be invariant under $G$ transformations. Under variations $\lambda$ of $U(1)^{(0)}$, the total fermion measure\footnote{In theories of gravity there would also be a gravitational anomaly $\sim \ \mathrm{tr} R \wedge R$ where $R$ is the Riemann curvature $2$-form.} then changes as
\begin{equation}
    D\Psi \rightarrow D\Psi \ \mathrm{exp} \left[- \frac{1}{16 \pi^{2}} \int d^{4}x \ \lambda \left(c_{a} F \widetilde{F} + c_{G}\mathrm{tr} G \widetilde{G}   \right)\right] \,.
\end{equation}
The constants $c_{a}$ and $c_{G}$ are determined by the exact fermion content. If the $i$-th Weyl-fermion $i$ sits in representations $(q_{i},r_{i})$ of $U(1)^{0} \times G$, then the total change of the measure is 
\begin{equation}
    c_{a} = \frac{1}{3} \sum_{i} q_{i}^{3} \mathrm{dim}(r_{i}) \,, \quad c_{G} = \sum_{i}q_{i} I(r_{i}) 
\end{equation}
where $I(r_{i})$ is the Dynkin index of representation $r_{i}$ normalized such that $I(\square) = 1$ for the fundamental representation. If the fermion content can be chosen to cancel the anomalous variations of \eqref{axionshiftanomaly}, then the gauge symmetry is non-anomalous. The instantons associated with a non-zero total instanton number $I = \frac{c_{a}}{16\pi^{2}} F \widetilde{F} + \frac{ c_{G}}{16 \pi^{2}} \mathrm{tr} G \widetilde{G}$ are then confined by the massless fermions and will only contribute to amplitudes involving these massless fermions in the initial/final states. 

One of the simplest examples is $c_{a} = 0$ and the gauge group $G = SU(2)$ with gauge field $B_{\mu}$ and field strength $G_{\mu \nu}$. The massive dual axion is coupled to the instantons as
\begin{equation}
    \mathcal{L} \supset \frac{F_{a}^{2}}{2} \left(\partial_{\mu} a - A_{\mu} \right)^{2} +  c_{G}\frac{a}{16 \pi^{2}} \mathrm{tr}G\widetilde{G} \,.
    \label{examplecouplinga}
\end{equation}
This theory is anomalous and requires the introduction of a set of four massless left-handed Weyl-fermions $\psi_{i}$ in the fundamental representation of $G$. We pick the charges of these fermions under $A^{\mu}$ to be the Fermat charges $q_{i} = \{3,4,5,-6\}$, such that the constants $c_{a} = 0$ and $c_{G} = 6$. The fermionic Lagrangian is
\begin{equation}
    \mathcal{L} \supset  \sum_{i}  \psi^{\dagger}_{i}  \overline{\sigma}_{\mu} \left( i\partial^{\mu} + B_{\mu} + q_{i} A_{\mu} \right) \psi_{i} \,.
    \label{system}
\end{equation}
The gauge field $A^{\mu}$ is coupled to the chiral current $j^{\mu}_{5}$, which satisfies the equations of motion
\begin{equation}
\label{chiraleom}
j^{\mu}_{5} =  \sum_{i}q_{i}\psi^{\dagger} \overline{\sigma}_{\mu} \psi \,, \quad 
\partial_{\mu} j^{\mu}_{5} =  \frac{c_{G}}{16 \pi^{2}}\mathrm{tr}G\widetilde{G} \,.
\end{equation}
The non-zero divergence implies that instantons act as a source for this chiral current. Every instanton vertex comes with associated fermion worldlines with a total non-zero chiral charge (see figure \ref{fig1}) and is confined. 

Classically, one can calculate the worldline contribution to the partition function of 
an instanton-anti-instanton dipole pair separated by a distance $R$ in time by studying the equations of motion \eqref{chiraleom}, which for such a configuration are
\begin{equation}
    \partial_{\mu} j^{\mu}_{5} =  \delta(x^{\mu}+\frac{R}{2} \hat{t}) - \delta(x^{\mu}-\frac{R}{2}\hat{t})   \,.
\end{equation}
 In a spherical coordinate system $(t,r,\phi,\theta)$, these equations are solved by
\begin{equation}
    j^{\mu}_{5} =  \delta(r) \mathrm{Rect}\left(\frac{t}{R}\right) \delta_{\mu,0} + \chi_{\mu} \,, 
\end{equation}
 where $\chi_{\mu}$ has zero divergence $\partial_{\mu} \chi^{\mu} = 0$ and represents the contribution of a compact particle worldline.  
 
 The first part of this current was already studied in \eqref{explicitcurrent} and integrating out the gauge field $A^{\mu}$ would lead to linear confinement, this time by chirally charged fermion zero modes. However, the vacuum contains massless fermions; the Wilson line interpretation of the fermion breaks down and quantum corrections may be large. A full calculation of the partition function would require both the fermion profile in an instanton-anti-instanton background and a control over these quantum corrections. We will not attempt this here. 

Instanton can become unconfined if the fermion zero mode Wilson lines attached to the instanton vertex can end in the bulk. For instance, an interaction with strength $g$ between the four fermions can be added of the form
\begin{equation}
    \mathcal{L} \supset - g_{1}  e^{-i c_{G} a} \left(\psi_{1} \psi_{2} \right) \left(\psi_{3}\psi_{4} \right) + \mathrm{h.c.}\,,
    \label{extramasss}
\end{equation}
or two two-point interactions
\begin{equation}
    \mathcal{L} \supset - g_{2}  e^{-i (q_{1}+q_{2}) a}\psi_{1} \psi_{2} -g_{3} e^{-i (q_{1}+q_{2}) a}\psi_{3}\psi_{4} + \mathrm{h.c.}
    \label{extramasss}
\end{equation}
The colour and Lorentz indices are contracted pair-wise using Levi-Civita symbols. 

The zero modes associated with the gauge-variant instanton vertex $e^{i c_{G}a}$ can end in the bulk by an insertion of the term above. The instanton no longer couples to the gauged axion and is unconfined. 

A second way to unconfine such instantons is the introduction of a second axion $b^{\mu}$, whose shift symmetry is also gauged,
\begin{equation}
     \mathcal{L} \supset \frac{F_{b}^{2}}{2} \left(\partial_{\mu} b - A_{\mu} \right)^{2} \,.
\end{equation}
Instantons can now be coupled to the gauge-invariant combination $a-b$ and become unconfined. This can be done either by directly altering \eqref{axionshiftanomaly} by $a\rightarrow a - b$ or adding a fermion mass of the form \eqref{extramasss} with $a \rightarrow b$. The gauge-invariant combination $a-b$ can then receive a potential in an instanton liquid. A second set of instantons coupled to the orthogonal gauge-variant combination $a+b$ would however still be confined. 

In the gauge $a = 0$, we recognize this simple toy model \eqref{system} as a massive vector field coupled to an anomalous current involving massless fermions. The dynamics confining the instantons in these simple models are equivalent to those of the massless up quark and its solution to the strong CP problem. We encourage further model-building efforts exhibiting instanton confinement beyond such simple toy models, particularly those in which the confining dynamics do not require the introduction of additional degrees of freedom.

The addition of the massive vector fields coupled to the non-conserved current can potentially lead to linear confinement due to the additional self-energy of the charged particles, but also leads to phenomenological signatures including amplitudes involving a longitudinal mode emission of the massive vector field proportional to the ratio (energy of the process/vector mass). For Standard Model currents, such couplings have been extensively studied in the literature (see \cite{Dror:2017nsg,Dror:2018wfl,Dror:2017ehi} and references therein) and include currents broken at tree level such as axial currents broken by fermion masses or tree-level conserved currents broken by a chiral anomaly, such as the SM baryon number or lepton number currents. Both couplings lead to FCNC and rare flavour-changing meson decays and can be constrained by missing energy signatures or visible decays depending on the specific model \cite{Dror:2017nsg}. By the Goldstone boson equivalence theorem, such couplings are equivalent to axion-like FCNC and constrained in \cite{Izaguirre:2016dfi}. The non-observation of such processes would imply that such couplings can only exist for small couplings $e$ or large masses $m$.

\section{Discussion}
We have studied a modification of axion physics in which the dual axion acquires a mass. In such a superconducting phase, axion strings are screened and far-separated instanton configurations are exponentially suppressed. Dipole configurations of instantons received an additional exponential suppression proportional to the distance between the individual instantons, corresponding to a charged particle travelling between the instantons. We have considered several models of such confinement with both dynamical instantons and dynamical worldlines. Much about the confinement of particles in non-supersymmetric theories is still poorly understood and we do not expect this to be an exhaustive list of instanton confinement models either.

The dynamics confining instantons and their contributions to the path integral in some of the simple models presented here were equivalent to the massless up quark and its solution to the strong CP problem. It would be interesting to further extend the list of such models to also include confinement of instantons akin to particle colour confinement in QCD in which the confining flux tubes are dynamically generated and do not require the introduction of additional degrees of freedom. Useful insight in such models of particle confinement has been gained in the past by considering supersymmetric QCD. In these theories, Seiberg and Witten \cite{Seiberg:1994rs} were able to demonstrate that, in a pure $SU(2)$ gauge theory, confinement of electric monopoles could be described by magnetic monopole condensation. Furthermore, by adding matter multiplets, they showed that it is possible to interpolate between a Higgs phase in which quarks are condensed to a phase in which magnetic monopoles condense \cite{Seiberg:1994aj}. It would be interesting to apply similar techniques to instanton confinement. Given the complexity of such an endeavor, we leave this as an open problem for future studies.

\section*{Acknowledgments}
I would like to thank Prateek Agrawal, Junwu Huang, Ethan Carragher, Matthew Reece and especially Mario Reig for helpful discussions. Any errors are my own. This paper was inspired by analogous considerations of a dual photon mass in the recent paper \cite{Hook:2022pcf}. Arthur Platschorre is supported by a STFC Studenship No. 2397217 and Cultuurfondsbeurs No. 40038041 made possible by the Pieter Beijer fonds and the Data-Piet fonds.

\bibliographystyle{utphys}
\bibliography{main}

\end{document}